\documentclass[10pt]{article}
\usepackage{arxiv}

\usepackage[]{graphicx}
\usepackage{newtxtext,newtxmath}

\usepackage[]{hyperref}
\usepackage{bookmark}
\hypersetup{
    setpagesize=false,
    bookmarksnumbered=true,
    bookmarksopen=true,
    colorlinks=true,
    linkcolor=black,
    citecolor=black,
    urlcolor=black,
}
\usepackage{xurl}

\usepackage[
    style=base,
    labelfont=bf,
    skip=6pt]{caption}
\usepackage{booktabs}

\usepackage{siunitx}
\def\figref#1{Fig.~\ref{#1}}
\def\tabref#1{Table~\ref{#1}}

\title{Analysis of Potential Generative AI Use in Abstracts of KAKENHI-Funded Projects}

\author{ 
    \href{https://orcid.org/0000-0002-4502-3164}
    {
        \includegraphics[scale=0.06]{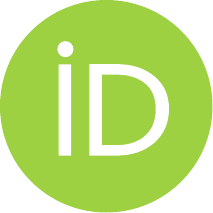}
        \hspace{1mm}
        Hitoshi KOSHIBA
    }\thanks{Corresponding.}\\
	\texttt{hkoshiba@ieee.org} \\
	\And
	\href{https://orcid.org/0009-0005-4830-4657}
    {
        \includegraphics[scale=0.06]{orcid.pdf}
        \hspace{1mm}
        Miki IDA-KIMURA
    } \\
}

\date{August 21, 2026}

\hypersetup{
    pdftitle={Analysis of Potential Generative AI Use in Abstracts of KAKENHI-Funded Projects},
    pdfsubject={cs.DL},
    pdfauthor={Hitoshi KOSHIBA, Miki IDA-KIMURA},
    pdfkeywords={KAKENHI, Funding, Grant, generative AI, LLM},
}

\begin{document}
%------------------------------------
\maketitle

%===========================================================
\begin{abstract}
This study analyzes the extent to which abstracts of projects funded under the Scientific Research (C) category of the Grants-in-Aid for Scientific Research (KAKENHI) were classified as AI-generated.

The analysis covers projects funded over the five-year period from FY2022 to FY2026.
The number of abstracts classified as AI-generated began to increase in FY2025,
and approximately \qty{20}{\percent} of the abstracts were classified as AI-generated in FY2026.
Although the proportions varied to some extent, abstracts classified as AI-generated were observed in many review categories.
These findings indicate that the use of generative AI has begun to spread across many research fields represented in Scientific Research (C).
\end{abstract}

% keywords can be removed
\keywords{KAKENHI \and Funding \and Grant \and generative AI \and LLM}

%===========================================================
\section{Introduction}
Since the official release of ChatGPT, a generative AI service, in November 2022,
the use of generative AI has spread widely throughout society.
Its use has also attracted attention in research settings, and the Japanese government is promoting initiatives related to AI for Science.

At the same time, it is difficult to investigate the actual use of generative AI in research.
One initial obstacle is the definition of ``generative AI use.''
Generative AI can support tasks at many different levels, including text summarization, translation, coding assistance, content review, and consultation when developing research questions.
Consequently, establishing a clear definition of generative AI use is itself a challenge.
Moreover, even after a definition has been established, external verification of actual use is often difficult, leaving researchers dependent on self-reporting.
Some academic journals have begun to require statements concerning the use of generative AI, but partly because of the definitional issues noted above, there is currently no standard format for such disclosures.

Given recent developments, it is unlikely that the use of generative AI will be generally prohibited or voluntarily avoided in future research practice.
Rather, it can be expected to become common in most research settings.
Nevertheless, the current situation remains transitional, and public policy is also actively promoting its use through initiatives such as AI for Science.
Under these circumstances, it is important to observe the extent to which generative AI use is spreading and the fields in which it is occurring.

Accordingly, with the aim of understanding the use of generative AI in research settings in Japan,
this study estimates and summarizes the proportion of abstracts classified as AI-generated among projects funded over the five-year period from FY2022 to FY2026 under the Grants-in-Aid for Scientific Research (KAKENHI), Scientific Research (C), administered by the Japan Society for the Promotion of Science (JSPS).

%----------------------
\subsection{Premises and Hypotheses}
The premises and hypotheses underlying the study are summarized below.

As noted above, both defining generative AI use and detecting it externally are difficult, and in most cases, the use of generative AI must therefore be determined based on individual researchers’ self-reports.
At present, self-reporting generally requires survey-based investigation, which places a substantial burden on both respondents and analysts and makes retrospective investigation difficult, such as attempting in FY2026 to determine patterns of use in FY2024.

This study therefore adopts an external estimation approach.
For the purposes of the analysis, generative AI use is defined simply as ``using generative AI to generate text.''
The analysis focuses on abstracts of funded KAKENHI for Scientific Research (C) projects.

The reasons for using external estimation are the cost of investigation and the difficulty of retrospective analysis described above.
The definition of use as ``text generation using generative AI'' is adopted for two reasons.
First, text is the only externally observable output available for the present analysis.
Even if most of the work on non-public experimental code were conducted with generative AI assistance, or if a research question were developed through dialogue with generative AI, such use would currently be difficult to detect externally.
Second, and related to the choice of analytical target, the study aims to observe the spread of generative AI use.
The use of generative AI to generate an abstract for a KAKENHI project is expected to be primarily a summarization task.
That is, researchers are likely to use generative AI to produce an abstract from a research plan they have already prepared.
This is a relatively low-barrier use case and may therefore be accessible even to researchers with limited experience using generative AI.

The decision to focus on KAKENHI Scientific Research (C) reflects its position as a relatively small-scale and broadly accessible funding category. 
With grants of up to JPY 5 million over three to five years, 
no restriction to a particular career stage,
and coverage across a wide range of disciplines, 
Scientific Research (C) attracts a large and diverse pool of applicants, from early-career to senior researchers.
In addition to temporal changes in the extent of use, differences among fields are important for this analysis.
Several constraints arise when attempting to study a broad range of fields within Japan.
For example, article abstracts could also be used for this purpose.
However, article datasets contain both English and Japanese texts, are difficult to restrict to Japan, and would be extremely large.
KAKENHI for Scientific Research (C), by contrast, consists primarily of Japanese-language texts, includes approximately \num{12000} projects per year, covers a broad range of fields, and is relatively easy to collect.
Policies concerning generative AI also vary among publishers of original research articles: some explicitly permit its use, whereas others have not established a policy, making it difficult to standardize conditions.
Furthermore, as noted above, generative AI use in producing project abstracts is expected to be relatively casual and is therefore well suited to the purpose of this study.

The point concerning a relatively low-barrier use case merits further explanation.
For many researchers, the ultimate objective is to publish research articles, while obtaining KAKENHI funding is only a means to that end.
Researchers can therefore be expected to have an incentive to reduce the cost of preparing applications whenever possible.
Taken together, these considerations suggest that the motivation and incentive to use generative AI in preparing project abstracts are substantial,
making such abstracts an appropriate target for capturing generative AI use across a wide range of researchers.

%===========================================================
\section{Method for Estimating AI-Generated Text}
The method used to estimate whether a text was generated by generative AI was adapted from the approach described in \cite{Liang2025}.

\begin{figure}[thb]
    \centering
    \includegraphics[width=\linewidth]{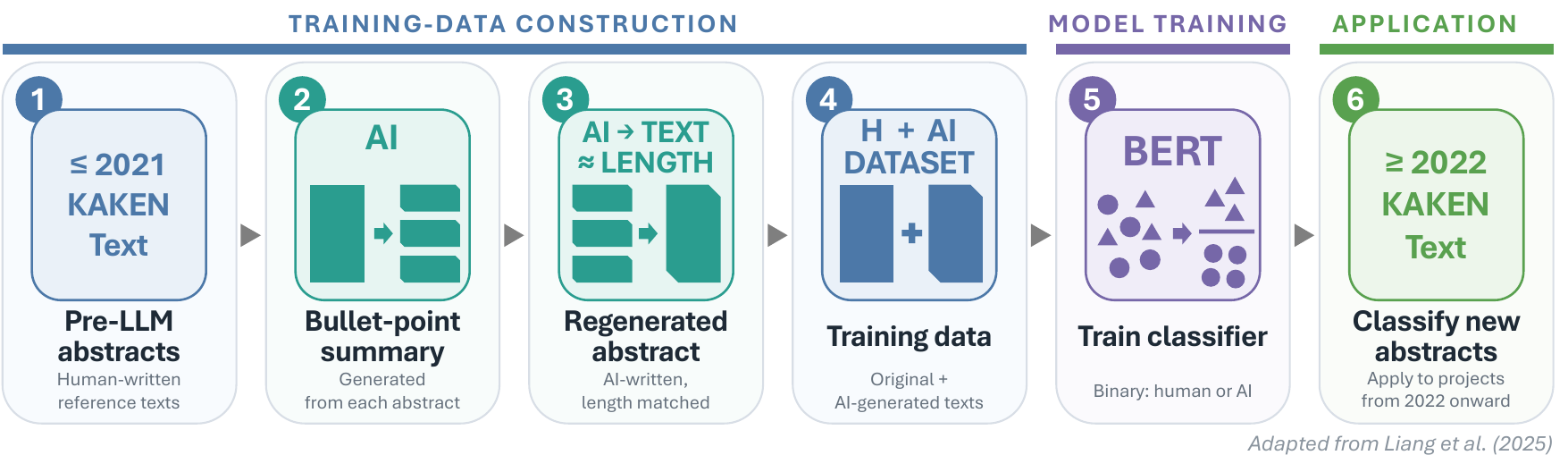}
    \caption{Workflow for detecting AI-generated abstracts}
    \label{fig:process}
\end{figure}

The basic procedure used in this study was as follows (\figref{fig:process}).
\begin{enumerate}
    \item Collect research projects from before the widespread adoption of generative AI (FY2021 or earlier).
    \item Use generative AI to convert each abstract into a bullet-point summary.
    \item Use the summary to generate an abstract of approximately the same length as the original.
    \item Prepare the original texts from Step 1 and the generated texts from Step 3 as training data.
    \item Train a binary BERT classifier for human-written versus AI-generated text.
    \item Use the classifier from Step 5 to estimate whether research project abstracts from FY2022 onward were human-written or AI-generated.
\end{enumerate}

The procedure from Step 5 onward differs substantially from that of \cite{Liang2025}.
The previous study estimated the mixture probability of human-written and AI-generated text by applying maximum likelihood estimation to differences in their word distributions.

By contrast, the present study uses BERT \cite{devlin-etal-2019-bert} for estimation.
This choice primarily reflects differences in the framework for performance validation and in the types of features considered.

The previous method does not estimate whether a text was written by a human or generated by an AI model.
Instead, it estimates the mixture proportion of human-written and AI-generated language in a collection of sentences.
Because users are unlikely to employ generative AI output without modification and are more likely to use it only in part, this conception is reasonable.

On the other hand, the previous method relies solely on word distributions and therefore discards contextual information, including the relationships among neighboring words.
It also requires preprocessing such as stop-word removal and handling of technical terms.
Because the present analysis primarily concerns Japanese text, language-specific differences in such preprocessing must also be considered separately.

Compared with methods based solely on word distributions, BERT can capture a wider variety of features, including contextual information. 
Its predictive performance can also be evaluated using standard supervised-learning frameworks. 
Moreover, multilingual adaptation is straightforward:
the same approach can be applied to different languages simply by replacing the pretrained model with a language-specific one, 
resulting in high versatility and reproducibility.

%===========================================================
\section{Data}\label{sec:data}
The analytical sample comprised funded KAKENHI for Scientific Research (C) projects from FY2018 through FY2026 for which an abstract was available.
The data were obtained in XML format through KAKEN\footnote{\url{https://kaken.nii.ac.jp/ja/}}.

For each project, the first element in the XML \texttt{paragraphList} was extracted.
This element corresponds to the ``Research Summary at the Start of the Project'' on KAKEN and is based on the ``Research Summary'' field in the grant application documents.

The four years from FY2018 through FY2021 were used as training data for constructing the AI-generated-text classifier,
while the five years from FY2022 through FY2026 constituted the actual analytical sample.
The number of records is shown in \tabref{tab:data_counts}.

\begin{table}[!htbp]
    \centering
    \caption{Number of records in the dataset}
    \label{tab:data_counts}
    {
    \begin{tabular}{cp{1mm}cp{1mm}r}
    \toprule
       \textbf{Type} && \textbf{Fiscal Year} && \textbf{Records} \\
    \midrule
        Training && 2018 && \num{11936}\\
        Training && 2019 && \num{12653}\\
        Training && 2020 && \num{12522}\\
        Training && 2021 && \num{12572}\\
    \midrule
        Analysis && 2022 && \num{12747}\\
        Analysis && 2023 && \num{11745}\\
        Analysis && 2024 && \num{12258}\\
        Analysis && 2025 && \num{12447}\\
        Analysis && 2026 && \num{12101}\\
    \bottomrule
    \end{tabular}
    }
\end{table}

%===========================================================
\section{Methods}\label{sec:methods}
\subsection{Generation of AI-Generated Text}

AI-generated texts were created using the procedure described above, with reference to the previous study \cite{Liang2025}.
Specifically, each original project abstract was first converted into a bullet-point summary,
after which a different LLM was used to regenerate an abstract from that summary in a two-stage procedure.
Rather than instructing an LLM to rewrite the original text directly,
the research content was first reduced to its essential points and then reconstructed as prose.
This procedure was intended to preserve as much information as possible about the original research content while allowing the wording itself to be newly generated by AI.

\subsubsection{Summarization of Research Project Abstracts}

In the first stage, each original project abstract was converted into a bullet-point list of nominalized phrases representing the principal points of the research.
Anthropic Claude Sonnet 4.5 was used for this process through AWS Bedrock
\footnote{\nolinkurl{jp.anthropic.claude-sonnet-4-5-20250929-v1:0}}.

\begin{figure}[thb]
    \centering
    \fbox{
    \includegraphics[width=0.84\linewidth]{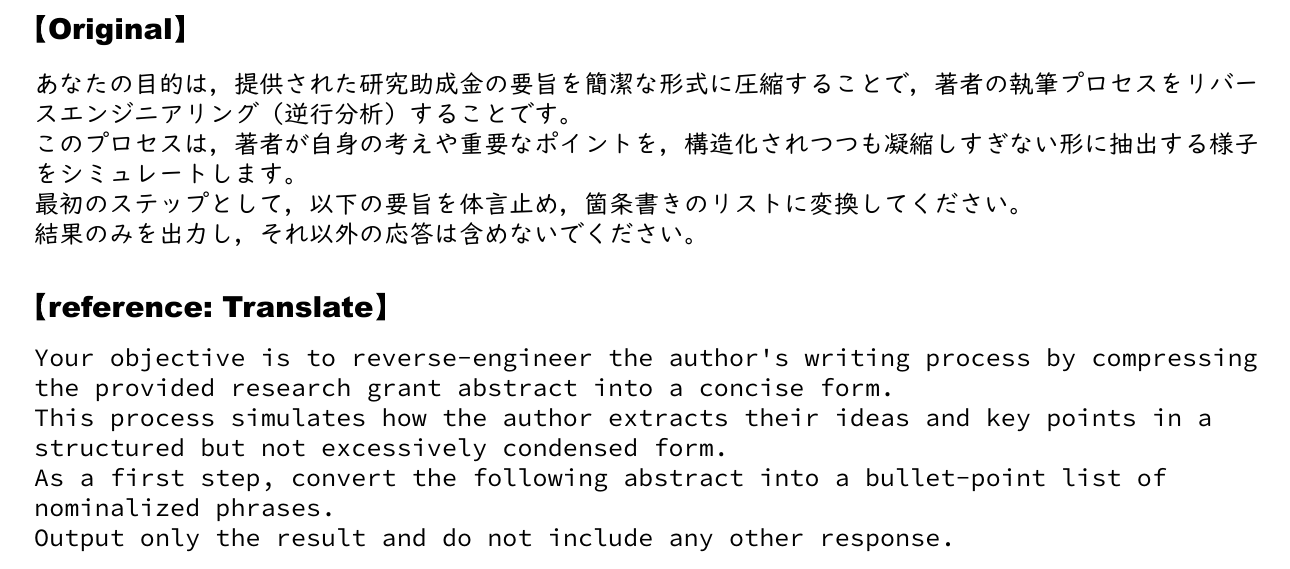}}
    \caption{Summarization prompt}
    \label{fig:prompt1}
\end{figure}

\figref{fig:prompt1} shows the summarization prompt.

For API calls, the generation settings were
\texttt{max\_tokens}=\num{65536} and \texttt{top\_p}=\num{1};
other generation parameters, including temperature, were not explicitly specified.

It was important for this summarization stage to extract the content appropriately.
A commercial LLM with strong text-understanding and summarization capabilities at the time of the study was therefore used.
Because the analysis concerned Japanese research project abstracts,
the prompts and processing procedure in \cite{Liang2025} were used as a reference while adapting the prompt to Japanese text.

\subsubsection{Regeneration of Research Project Abstracts}

\begin{figure}[thb]
    \centering
    \fbox{
    \includegraphics[width=0.84\linewidth]{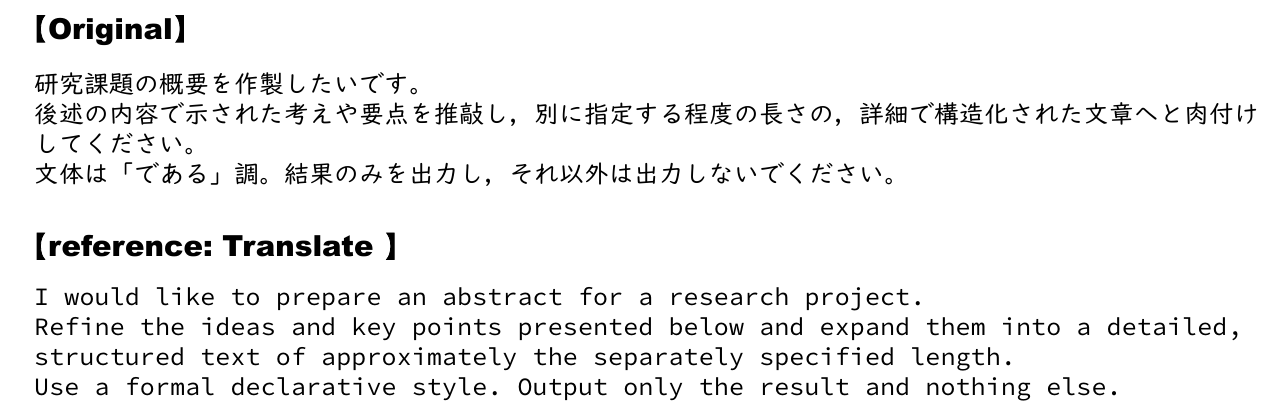}}
    \caption{Abstract generation prompt}
    \label{fig:prompt2}
\end{figure}

In the second stage, new project abstracts were generated from the bullet-point summaries obtained in the first stage.
\figref{fig:prompt2} shows the prompt used for abstract generation.

In addition to the bullet-point summary, the LLM was given the character count of the original project abstract
and instructed to generate a text of approximately the same length.
This was intended to minimize the possibility that text length itself would become a feature distinguishing human-written from AI-generated text because of a substantial difference between the lengths of the generated and original texts.

Local LLMs, rather than commercial LLMs, were used to generate the abstracts.
Generative AI capabilities improve rapidly.
If the training texts had been generated only with high-performance models available in 2026,
they might have exhibited characteristics different from those of AI-generated texts produced by models used in earlier years.
Commercial LLMs are also subject to model discontinuation and specification changes, making it impossible to guarantee continued access to the same model.
Local LLMs were therefore used so that the model versions could be held fixed and run repeatedly under reproducible conditions.

Multiple local LLMs were considered to avoid excessive dependence on expressions specific to any one model.

As a preliminary assessment, generation was tested with the following six models.
\begin{itemize}
    \item Qwen3 32B\footnote{The model designation in Ollama was \nolinkurl{qwen3:32b}.}~\cite{qwen3technicalreport}
    \item Llama 3.1 Swallow 8B\footnote{\nolinkurl{hf.co/mmnga/Llama-3.1-Swallow-8B-Instruct-v0.5-gguf}}~\cite{Fujii:COLM2024,Okazaki:COLM2024,ma:arxiv2025}
    \item gpt-oss-20b\footnote{The model designation in Ollama was \nolinkurl{gpt-oss:20b}.}~\cite{openai2025gptoss}
    \item Llama 3.3 Swallow 70B\footnote{\nolinkurl{hf.co/mmnga/tokyotech-llm-Llama-3.3-Swallow-70B-Instruct-v0.4-gguf:Q3_K_M}}~\cite{Fujii:COLM2024,Okazaki:COLM2024,ma:arxiv2025}
    \item LLM-jp-4 8B Thinking\footnote{\texttt{hf.co/mmnga-o/llm-jp-4-8b-thinking-gguf}}
    \item ELYZA Japanese Llama 2 13B\footnote{\nolinkurl{hf.co/second-state/ELYZA-japanese-Llama-2-13b-fast-instruct-GGUF:Q4_K_M}}~\cite{elyzallama2023}
\end{itemize}

This comparison was not intended as a quantitative benchmark of model performance.
The practical suitability of the models was instead assessed with emphasis on their ability to consistently generate the texts required for the present analysis,
using generation speed, instruction following, and the absence of obvious breakdowns in the generated Japanese prose as criteria.
Some candidate models required an extremely long time to generate text,
failed to follow the specified format or content,
or produced unnatural Japanese.

Based on this assessment,
``Qwen3 32B'' and ``Llama 3.1 Swallow 8B'' were ultimately selected.
Each model independently received the same bullet-point summary and the character count of the original abstract.
Generating texts with multiple models was intended to reduce the risk that the resulting classifier would depend only on vocabulary or expressions specific to a particular generation model.

The local LLMs were run using \texttt{Ollama 0.22.0}.
``Qwen3 32B'' was executed by specifying the corresponding model name in Ollama.
For API calls,
\texttt{num\_predict}=\num{5000} and \texttt{top\_p}=\num{1}
were specified; other generation parameters were not explicitly specified.
Quantized models in GGUF format were used for some candidates.
The local LLMs were run on a computer equipped with an \texttt{NVIDIA RTX A6000} GPU.

\subsection{Construction of a BERT Classifier for AI-Generated Text}

A binary BERT classifier distinguishing human-written from AI-generated text was constructed using the AI-generated texts produced by the procedure above and their corresponding human-written originals.
The base model was the pretrained Japanese BERT model developed by the Tohoku University Natural Language Processing Group
\footnote{\nolinkurl{cl-tohoku/bert-base-japanese-v3}}.

During training,
human-written texts were labeled \texttt{human},
and AI-generated texts were labeled \texttt{llm}.
Only the text itself was supplied to BERT.
Additional information, such as the name of the LLM used to generate a text, was not included.
The classifier therefore distinguished human-written from AI-generated text on the basis of features contained in the text itself.

A basic tokenizer was used for text processing.
The maximum sequence length was 512 tokens, and text beyond this limit was truncated from the end.
Texts shorter than 512 tokens were dynamically padded to the length of the longest sequence in each minibatch.

A sequence-classification output layer with two classes, \texttt{human} and \texttt{llm}, was added to the pretrained BERT model,
and the entire model was fine-tuned on the training data.
Training was conducted with \num{3} epochs,
a batch size of \num{16},
a learning rate of $2\times10^{-5}$,
and weight decay of \num{0.01}.
The first \qty{10}{\percent} of the total training steps was used as a warm-up period,
the number of gradient accumulation steps was set to 1,
and the random seed was \num{42}.

Training was conducted using Hugging Face Transformers.
This procedure produced a binary classification model that receives a research project abstract and predicts whether it is human-written or AI-generated.

%===========================================================
\section{Experiments}
\subsection{Generation of the Classifier}
The classifier was generated and trained using the data described in Section~\ref{sec:data} and the procedure described in Section~\ref{sec:methods}.

The initial training set consisted of \num{49683} Scientific Research (C) projects funded from FY2018 through FY2021 for which abstracts were available.

Bullet-point summaries were first generated for these projects.
Claude declined to process content related to viruses and other information with potential dual-use or weaponization concerns, reducing the number of successfully generated summaries by approximately 200 to \num{49481}.

Abstracts were then generated from these summaries using
``Qwen3 32B'' and ``Llama 3.1 Swallow 8B.''

The resulting training dataset consisted of \num{49683} human-written records
and \num{49481} AI-generated records from each of the two models, for a total of \num{98962} AI-generated records
and \num{148645} records overall.

\subsection{Evaluation of the Classifier}

\begin{table*}[htb]
    \centering
    \caption{Five-fold cross-validation results for the BERT classifier}
    \label{tab:bert_cv}
    \begin{tabular}{lrrrrrrr}
        \toprule
        & \textbf{Accuracy}
        & \multicolumn{3}{c}{\textbf{Macro}}
        & \multicolumn{3}{c}{\textbf{LLM class}} \\
        \cmidrule(lr){3-5}
        \cmidrule(lr){6-8}
        Fold
        &
        & Precision
        & Recall
        & F1
        & Precision
        & Recall
        & F1 \\
        \midrule
        1 & \num{0.9974} & \num{0.9971} & \num{0.9970} & \num{0.9971} & \num{0.9980} & \num{0.9981} & \num{0.9980} \\
        2 & \num{0.9972} & \num{0.9971} & \num{0.9966} & \num{0.9969} & \num{0.9973} & \num{0.9985} & \num{0.9979} \\
        3 & \num{0.9975} & \num{0.9974} & \num{0.9970} & \num{0.9972} & \num{0.9978} & \num{0.9985} & \num{0.9981} \\
        4 & \num{0.9971} & \num{0.9971} & \num{0.9963} & \num{0.9967} & \num{0.9970} & \num{0.9986} & \num{0.9978} \\
        5 & \num{0.9965} & \num{0.9968} & \num{0.9953} & \num{0.9960} & \num{0.9959} & \num{0.9988} & \num{0.9974} \\
        \midrule
        Mean
          & \num{0.9971} & \num{0.9971} & \num{0.9964} & \num{0.9968}
          & \num{0.9972} & \num{0.9985} & \num{0.9978} \\
        SD
          & \num{0.0004} & \num{0.0002} & \num{0.0007} & \num{0.0005}
          & \num{0.0008} & \num{0.0003} & \num{0.0003} \\
        OOF
          & \num{0.9971} & \num{0.9971} & \num{0.9964} & \num{0.9968}
          & \num{0.9972} & \num{0.9985} & \num{0.9978} \\
        \bottomrule
    \end{tabular}
\end{table*}

\begin{table}[htb]
    \centering
    \caption{Confusion matrix for out-of-fold predictions}
    \label{tab:bert_cv_confusion}
    \begin{tabular}{lrrr}
        \toprule
        & \multicolumn{2}{c}{\textbf{Predicted}} & \\
        \cmidrule(lr){2-3}
        \textbf{Actual} & Human & LLM & Total \\
        \midrule
        Human & \num{49405} &   \num{278} &  \num{49683} \\
        LLM   &   \num{149} & \num{98813} &  \num{98962} \\
        \midrule
        Total & \num{49554} & \num{99091} & \num{148645} \\
        \bottomrule
    \end{tabular}
\end{table}
Classifier performance was evaluated using stratified group $k$-fold cross-validation with $k=5$.

Stratification was used to keep the proportions of human-written (\texttt{human}) and AI-generated (\texttt{llm}) records as even as possible across folds, and the ID of the original research project was used as the grouping variable.
In principle, two AI-generated texts were created for each original human-written abstract
\footnote{In cases where Claude declined to generate the summary, only the original human-written abstract was available.}.
If these records were divided into training and validation sets by ordinary random splitting, a human-written text and one or more AI-generated texts with the same research content could be placed on opposite sides of the split.
The classifier might then exploit similarity in research content rather than features associated with the source of the writing, leading to an overestimation of classification performance.
To avoid this problem, all texts derived from the same original research project were assigned to the same fold.
For example, when a project had an original human-written text, a Qwen3-generated text, and a Swallow-generated text, all three were placed in the same fold.
Consequently, no text with the same research content as a validation record was included in the training data.

For each fold, fine-tuning started independently from the pretrained model; weights learned in a preceding fold were not carried over.
Five-fold cross-validation produced an out-of-fold (OOF) prediction for every text from a model that had not been trained on that text.
Performance was evaluated using Accuracy, Macro Precision, Macro Recall, and Macro F1.
Because detecting AI-generated text was of particular importance in this study,
Precision, Recall, and F1 were also calculated with \texttt{llm} as the positive class.
A confusion matrix consisting of True Negatives, False Positives, False Negatives, and True Positives was also calculated to examine cases in which human-written texts were misclassified as AI-generated and vice versa.

The results are shown in \tabref{tab:bert_cv} and \tabref{tab:bert_cv_confusion}.
Both Accuracy and Macro Recall were approximately \qty{99.7}{\percent}, indicating extremely high performance.
\tabref{tab:bert_cv_confusion} shows that, in relative terms, human-written texts were somewhat more likely to be misclassified as AI-generated; however, the absolute number was only approximately \num{300} out of \num{50000}.

It should be noted that \cite{Liang2025} processes texts at the sentence level, with a sentence extending to the Japanese full stop, whereas the present study processes the entire abstract as a single unit because of differences in methodology
\footnote{Sentence-level classification is also possible}.

After classifier performance had been confirmed by cross-validation, a separate final model was constructed for application to the actual project abstracts.

The final model did not reuse any of the fold-specific models.
Instead, fine-tuning was restarted from the pretrained BERT model using the entire training dataset under the same training conditions.
This model trained on all available data was used for subsequent estimation of AI-generated text.

At inference time, only the text of the target project abstract was supplied to the trained model.
The input was tokenized in the same manner as during training and limited to 512 tokens.
A softmax function was applied to the two class logits produced by the model to obtain a score for human-written text and a score for AI-generated text.
The class with the higher score was assigned as the predicted label.

\subsection{Classification of Project Abstracts}
\begin{figure}
    \centering
    \includegraphics[width=0.6\linewidth]{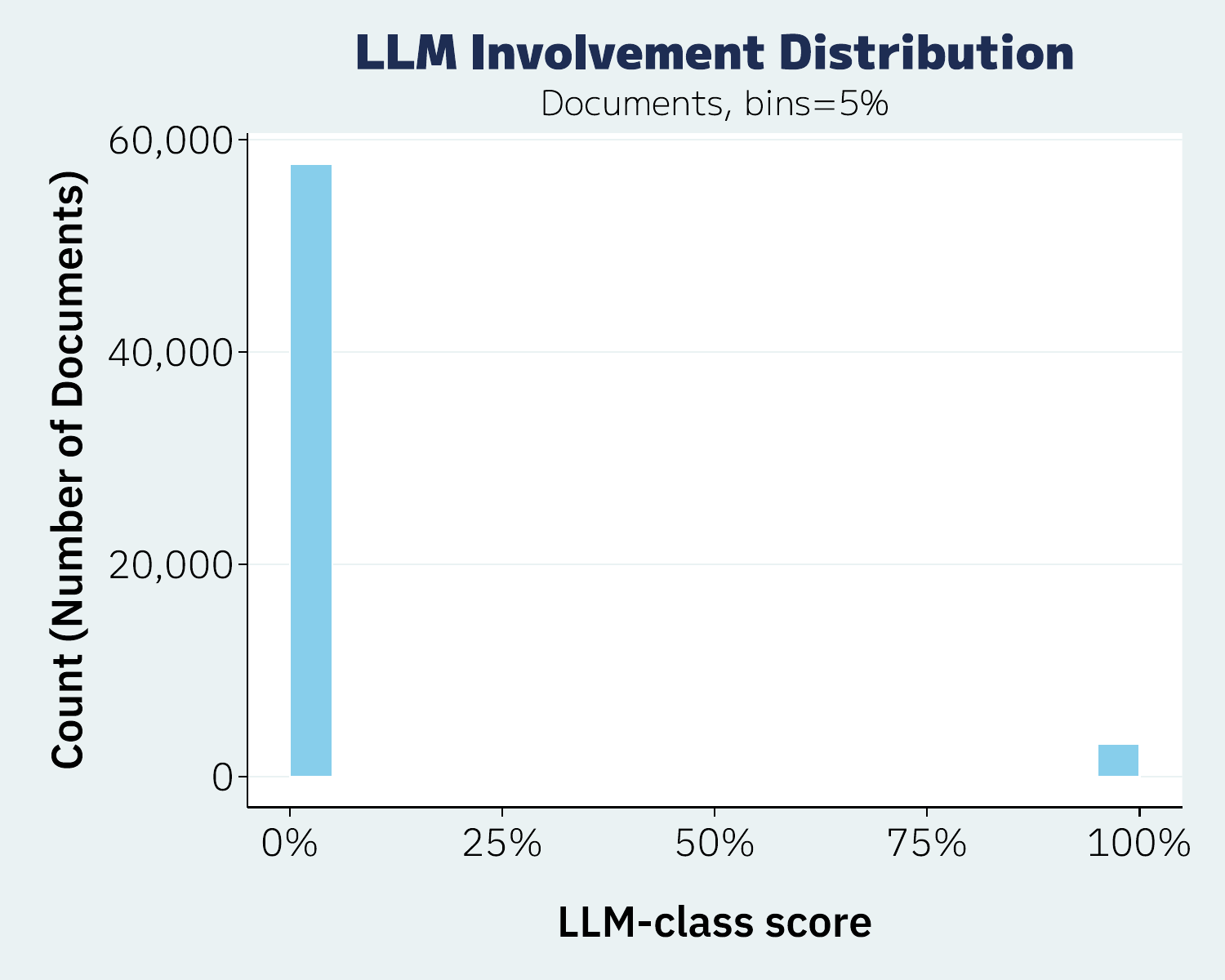}
    \caption{Distribution of estimated AI-generation scores}
    \label{fig:hist_doc}
\end{figure}

The classifier was applied to project abstracts from FY2022 onward to estimate whether they resembled AI-generated text.
The distribution of estimated AI-generation scores is shown in
\figref{fig:hist_doc}.

The distribution shows that cases with intermediate AI-generation scores, such as around \qty{70}{\percent}, were rare. Instead, scores were concentrated near \qty{0}{\percent} and \qty{100}{\percent}.

\subsubsection{Annual Trends}
\begin{figure}
    \centering
    \includegraphics[width=0.6\linewidth]{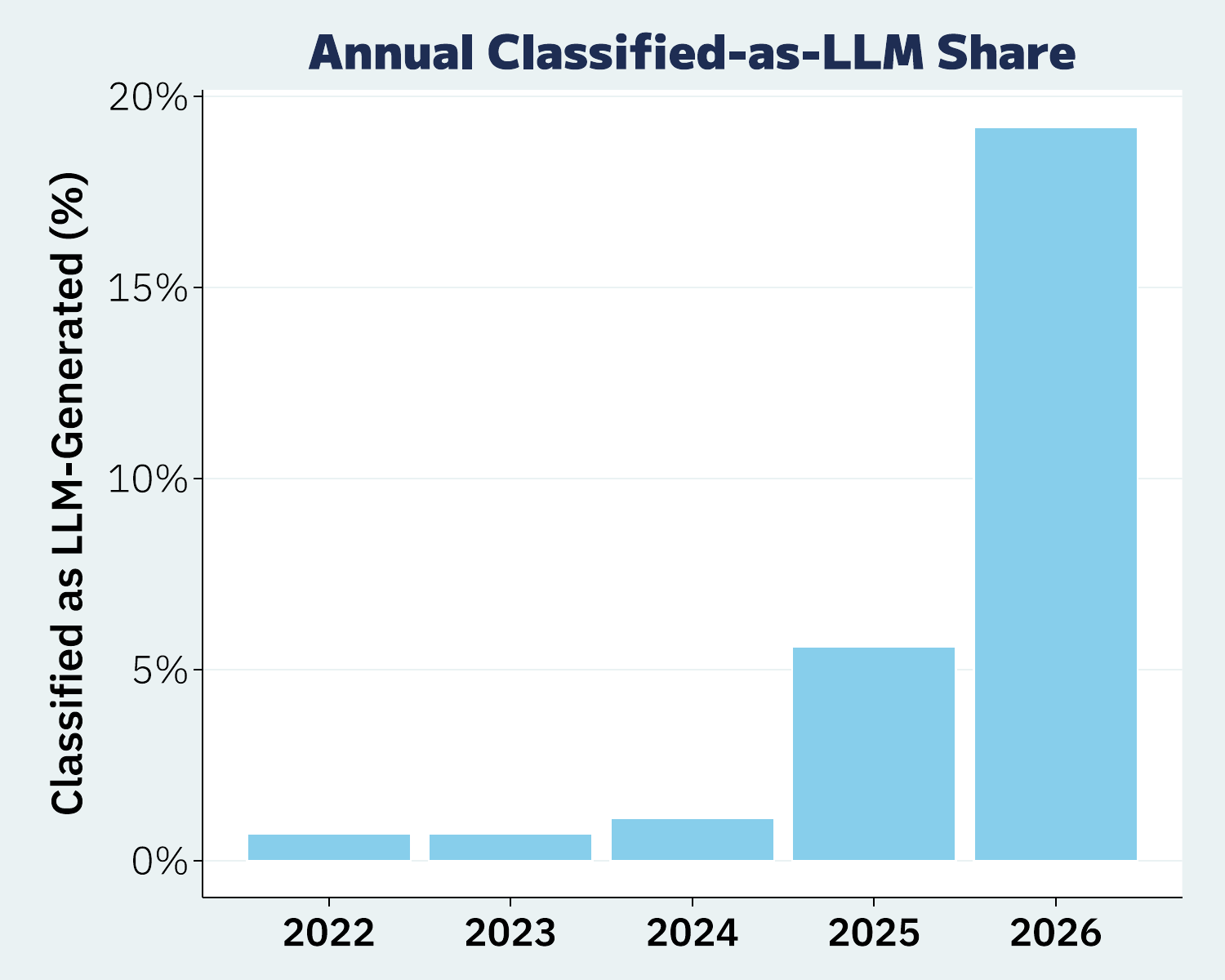}
    \caption{Annual trend in the percentage of project abstracts estimated to be AI-generated}
    \label{fig:series}
\end{figure}

The annual trend is shown in \figref{fig:series} and \tabref{tab:tbl_whole_series}.
These figures and tables indicate that abstracts estimated to involve generative AI became clearly observable in FY2025 and increased sharply in FY2026 to approximately \qty{20}{\percent}.

\begin{table}[tb]
    \centering
    \caption{Annual trend in the number of project abstracts estimated to be AI-generated}
    \label{tab:tbl_whole_series}
    \begin{tabular}{cp{2mm}rp{2mm}rp{2mm}r}
        \toprule
        \textbf{Fiscal Year} && \textbf{Total} && \textbf{LLM} && \textbf{PCT}\\
        \midrule
        2022 && \num{12747}&&   \num{92} &&  \qty{0.7}{\percent}\\
        2023 && \num{11745}&&   \num{88} &&  \qty{0.7}{\percent}\\
        2024 && \num{12258}&&  \num{134} &&  \qty{1.1}{\percent}\\
        2025 && \num{12447}&&  \num{696} &&  \qty{5.6}{\percent}\\
        2026 && \num{12101}&& \num{2321} && \qty{19.2}{\percent}\\
        \bottomrule
    \end{tabular}
\end{table}

ChatGPT was officially released in November 2022, and 
the deadline for the relevant post-award procedures for Scientific Research (C) is generally around April of the award year.
Accordingly, the abstracts for projects funded in FY2022 can be assumed to have been written by around April 2022,
and those for projects funded in FY2023 around April 2023.
It is therefore natural that the estimated proportion for FY2022 was close to zero, and records detected as AI-generated in that year are more likely to be false positives.
In FY2023, less than half a year had passed since the launch of ChatGPT.
Given the capabilities of the models available at the time, even when generative AI was used, substantial human revision was likely to have been required.
An increase beginning around FY2024 is therefore plausible.

Although direct comparison is not possible because the methods and other conditions differ,
this result is also consistent with a previous study of the U.S. NSF and NIH \cite{Qian2026}, 
which reported estimates of LLM-modified text reaching approximately \qty{20}{\percent} in some datasets.

To examine the robustness of the results to the choice of estimation method, LLM involvement was also estimated using word-occurrence distributions and maximum likelihood estimation, as in the previous study (Appendix~\ref{apdx:individual_abstract_results}).
Although the absolute level of the estimates obtained with that method differed from the binary classification results produced by BERT, both methods showed a common temporal pattern: estimates were broadly stable from FY2022 through FY2024, began to rise in FY2025, and increased substantially in FY2026.
This suggests that the increase observed from FY2025 onward is not attributable solely to characteristics specific to the BERT classifier.

\subsubsection{Trends by Field}
\begin{figure}
    \centering
    \includegraphics[width=0.6\linewidth]{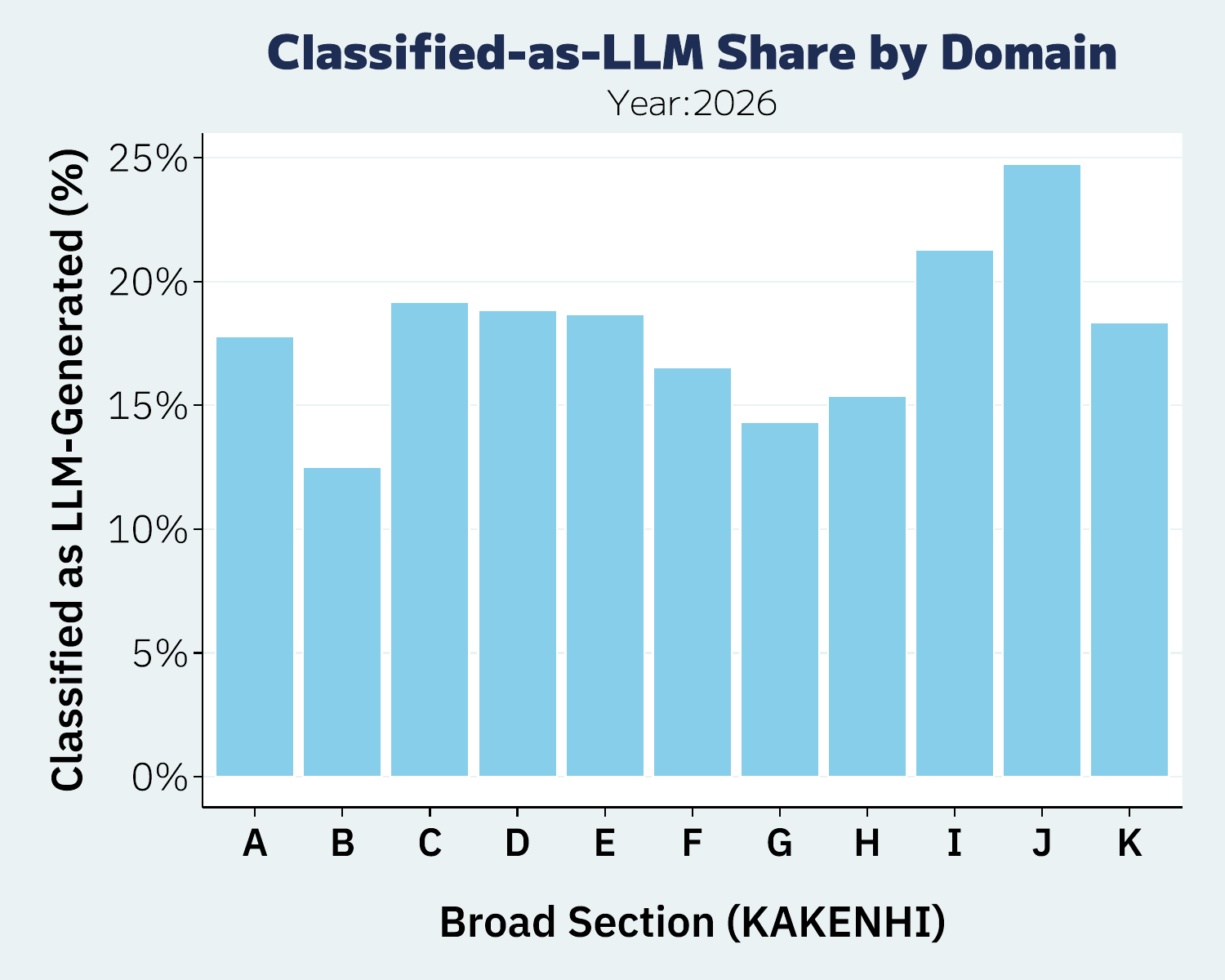}
    \caption{Percentage of abstracts estimated to be AI-generated by Broad Section}
    \label{fig:domain}
\end{figure}

For 2026, the classification results were aggregated by the broad sections used in the KAKENHI review-category system, as shown in \figref{fig:domain} and \tabref{tab:tbl_domain}.

Each project is assigned a Basic Section as its review category.
Basic Sections are generally organized hierarchically under Medium-sized Sections and Broad Sections; this correspondence was used to assign each project to a Broad Section.
However, Basic Sections belonging to Medium-sized Sections 80 and 90 correspond to multiple Broad Sections and therefore cannot be assigned uniquely.
Projects in Medium-sized Sections 80 and 90 were consequently excluded from this analysis.
The Broad Section aggregates reported below therefore do not constitute a complete and strictly exhaustive classification of all projects.

\begin{table}[tb]
    \centering
    \caption{Number of abstracts estimated to be AI-generated by Broad Section}
    \label{tab:tbl_domain}
    \begin{tabular}{cp{2mm}lp{2mm}rp{2mm}rp{2mm}r}
        \toprule
        \begin{tabular}{c}
        \textbf{\small Broad}\\\textbf{\small Section}
        \end{tabular}
         && 
         \textbf{Provisional English Label} &&
         \textbf{Total} && \textbf{LLM} && \textbf{PCT}\\
        \midrule
        A && {\small Humanities and Social Sciences}&& \num{3235} && \num{575} &&  \qty{17.8}{\percent}\\
        B && {\small Mathematics and Physical Sciences}&& \num{527} &&  \num{66} &&  \qty{12.5}{\percent}\\
        C && {\small Engineering}&& \num{710} && \num{136} &&  \qty{19.2}{\percent}\\
        D && {\small Materials and Applied Engineering}&& \num{260} &&  \num{49} &&  \qty{18.8}{\percent}\\
        E && {\small Chemistry}&&\num{289} &&  \num{54} &&  \qty{18.7}{\percent}\\
        F && {\small Agricultural Sciences}&& \num{611} && \num{101} &&  \qty{16.5}{\percent}\\
        G && {\small Biological Sciences and Neuroscience}&& \num{349} &&  \num{50} &&  \qty{14.3}{\percent}\\
        H && {\small Pharmaceutical and Basic Medical Sciences}&& \num{514} &&  \num{79} &&  \qty{15.4}{\percent}\\
        I && {\small Medical and Health Sciences}&&\num{4565} && \num{971} &&  \qty{21.3}{\percent}\\
        J && {\small Informatics}&& \num{408} && \num{101} &&  \qty{24.8}{\percent}\\
        K && {\small Environmental Sciences}&& \num{158} &&  \num{29} &&  \qty{18.4}{\percent}\\
        \bottomrule
         \multicolumn{9}{r}{
         \tiny 
        Aggregated across sections, excluding items in Medium-sized Sections 80 and 90.
         }\\
    \end{tabular}
\end{table}

\figref{fig:domain} and \tabref{tab:tbl_domain} show that Broad Section J had the highest proportion.
Broad Section J includes Medium-sized Sections related to informatics and is therefore expected to include many researchers familiar with generative AI technology, making this result plausible.
Broad Section B, corresponding to mathematical and physical sciences, had the lowest proportion.

The ten Medium-sized Sections with the highest and lowest proportions are shown in
\tabref{tab:top10} and \tabref{tab:bottom10}.
These tables indicate that substantial field-level differences are present at the Medium-sized Section level.

\begin{table}[tb]
    \centering
    \caption{Medium-sized Sections with the highest proportions of AI-generated abstracts: top 10}
    \label{tab:top10}
    \small
    \begin{tabular}{clrrr}
        \toprule
        \begin{tabular}{c}
            \textbf{\scriptsize Section}\\
            \textbf{\scriptsize Broad-Mid}
        \end{tabular}&
        \textbf{Label} &
        \textbf{Total}  &
        \textbf{LLM}  &
        \textbf{PCT} \\
        \midrule
        J - 62 & Applied informatics and related fields & \num{91} & \num{29}& \qty{31.9}{\percent}\\
        J - 61 & Human informatics and related fields & \num{161} & \num{47}& \qty{29.2}{\percent}\\
        C - 22 & Civil engineering and related fields & \num{115} & \num{30}& \qty{26.1}{\percent}\\
        I - 51 & Brain sciences and related fields & \num{51} & \num{13}& \qty{25.5}{\percent}\\
        I - 59 & {\footnotesize Sports sciences, physical education, health sciences, and related fields} & \num{623} & \num{154}& \qty{24.7}{\percent}\\
        A -  7 & Economics, business administration, and related fields & \num{413} & \num{100}& \qty{24.2}{\percent}\\
        D - 31 & {\scriptsize Nuclear engineering, earth resources engineering, energy engineering, and related fields} & \num{29} & \num{7}& \qty{24.1}{\percent}\\
        K - 64 & Environmental conservation measure and related fields & \num{99} & \num{23}& \qty{23.2}{\percent}\\
        E - 36 & {\small Inorganic materials chemistry, energy-related chemistry, and related fields} & \num{35} & \num{8}& \qty{22.9}{\percent}\\
        E - 35 & Polymers, organic materials, and related fields & \num{44} & \num{10}& \qty{22.7}{\percent}\\
        \bottomrule
    \end{tabular}
\end{table}

\begin{table}[tb]
    \centering
    \caption{Medium-sized Sections with the lowest proportions of AI-generated abstracts: bottom 10}
    \label{tab:bottom10}
    \small
    \begin{tabular}{clrrr}
        \toprule
        \begin{tabular}{c}
            \textbf{\scriptsize Section}\\
            \textbf{\scriptsize Broad-Mid}
        \end{tabular}&
        \textbf{Label} &
        \textbf{Total} &
        \textbf{LLM} &
        \textbf{PCT} \\
        \midrule
        A - 4 & Geography, cultural anthropology, folklore, and related fields & \num{89}& \num{4} & \qty{4.5}{\percent}\\
        B - 11 & Algebra, geometry, and related fields & \num{100}& \num{5} & \qty{5.0}{\percent}\\
        A -  5 & Law and related fields & \num{180}& \num{13} & \qty{7.2}{\percent}\\
        A -  3 & History, archaeology, museology, and related fields & \num{210}& \num{18} & \qty{8.6}{\percent}\\
        E - 33 & Organic chemistry and related fields & \num{58}& \num{5} & \qty{8.6}{\percent}\\
        G - 46 & Neuroscience and related fields & \num{60}& \num{6} & \qty{10.0}{\percent}\\
        K - 63 & Environmental analyses and evaluation and related fields & \num{59}& \num{6} & \qty{10.2}{\percent}\\
        B - 14 & Plasma science and related fields & \num{19}& \num{2} & \qty{10.5}{\percent}\\
        G - 44 & Biology at cellular to organismal levels, and related fields & \num{103}& \num{11} & \qty{10.7}{\percent}\\
        C - 24 & {\small Aerospace engineering, marine and maritime engineering, and related fields} & \num{36}& \num{4} & \qty{11.1}{\percent}\\
        \bottomrule
    \end{tabular}
\end{table}

\subsubsection{Results by Academic Position (Proxy for Age)}
Although KAKEN does not provide the ages of funded researchers, it provides information on academic positions such as professor and assistant professor.
Position was therefore used as a rough proxy for age, and projects classified as AI-generated from fiscal years FY2022 through FY2026 were aggregated by position.
The results are shown in \tabref{tab:job_title}.

\begin{table}[htb]
    \centering
    \caption{Academic positions among projects estimated to be AI-generated: top five categories}
    \label{tab:job_title}
    \begin{tabular}{lp{2mm}rp{1mm}rp{1mm}r}
        \toprule
        \textbf{Position} && 
        \textbf{Total} &&
        \textbf{LLM} &&
        \textbf{PCT} \\
        \midrule
        Professor           && \num{20096} &&  \num{890} && \qty{4.4}{\percent}\\
        Associate Professor && \num{14933} &&  \num{780} && \qty{5.2}{\percent}\\
        Assistant Professor &&  \num{7478} &&  \num{517} && \qty{6.9}{\percent}\\
        Lecturer            &&  \num{6191} &&  \num{384} && \qty{6.2}{\percent}\\
        Researcher          &&  \num{1475} &&   \num{89} && \qty{6.0}{\percent}\\
        \bottomrule
    \end{tabular}
\end{table}

Because position titles vary widely, they were not consolidated,
and only the five most frequent categories are reported in \tabref{tab:job_title}.
The pattern could differ if variant titles (e.g., specially appointed professor, associate professor, or assistant professor) were consolidated with their corresponding standard titles.
In the results shown above, professor was the most frequent category, followed by associate professor and assistant professor.
Because professors, associate professors, and assistant professors can generally be expected to be ordered from older to younger,
these results suggest that relatively older researchers may also be using generative AI.

%===========================================================
\section{Conclusion}
With the aim of understanding generative AI use in research in Japan,
this study constructed a classifier to classify project abstracts as human-written or AI-generated and examined the proportion classified as AI-generated.

Approximately \qty{20}{\percent} of the abstracts for projects funded in FY2026 were estimated to be AI-generated.
Differences among Broad Sections were not large: the proportion was approximately \qty{25}{\percent} in informatics-related fields and approximately \qty{10}{\percent} even in mathematical and physical sciences, where it was lowest.
At the Medium-sized Section level, however, the proportions ranged widely, from over \qty{30}{\percent} to below \qty{5}{\percent}, and this variation warrants caution.
Projects classified as AI-generated were observed among professors, associate professors, and assistant professors. This suggests that the use of generative AI may not be confined to younger researchers.

Although direct comparison is difficult because of differences in methodology and units of analysis, a previous study of U.S. NSF and NIH grant proposals and awards \cite{Qian2026} likewise found substantial LLM involvement in grant texts. Its corpus-level estimates of the fraction of LLM-modified sentences rose sharply after 2023, reaching approximately \qtyrange{8}{20}{\percent} across the four datasets by 2025. These estimates are of a similar order of magnitude to the present estimate of ``approximately \qty{20}{\percent} of projects.''

These findings suggest that the magnitude of LLM involvement observed in Japanese KAKENHI project abstracts is broadly comparable to that reported for U.S. federal research funding texts, although the estimates are not directly comparable.

\subsection{Limitations and Future Research}
Beginning with the fiscal year 2024 KAKENHI call for proposals, released in July 2023, the application guidelines included provisions concerning generative AI.
Use itself is not prohibited; researchers are instructed to make their own decisions while taking the various risks into account.
The authors also do not regard the use of generative AI to prepare research project abstracts as inherently problematic.
Even when an abstract is classified as AI-generated, whether that particular abstract was actually generated or summarized by generative AI remains a separate question.
Indeed, misclassification occurred even in the controlled training environment designed to exclude AI-generated material from the human-written data.
In the preliminary experiments, several local LLMs also produced outputs that did not follow the instructions.
In other words, capabilities differ among LLMs.
The present study was limited to two specific models, Qwen3 and Swallow.
The stability of the results when switching to other LLMs remains to be examined.

Relatedly, generative AI and LLMs are evolving rapidly.
Since the release of ChatGPT in November 2022, new models have appeared in succession and their capabilities have continually improved.
Accordingly, the performance and characteristics of the LLMs used in 2023 may differ substantially from those used in 2026,
and the relationship between time period and model type warrants further study.

The present analysis is limited to funded Scientific Research (C) projects and therefore does not necessarily represent generative AI use in other KAKENHI categories or in Japanese research activities more broadly.

%===========================================================

%===========================================================
\clearpage
\appendix
{\LARGE\textbf{Appendix}}
\section{Sentence-Level Estimation}
A previous study \cite{Qian2026} examined the proportion of AI-generated text in U.S. NSF and NIH funding proposals.
Although the study is closely related to the present work, direct comparison is difficult because of differences in methodology, including its reliance on word frequencies, estimation of mixture proportions, and use of sentences rather than whole abstracts as the analytical unit, as well as differences in language and funding structures.

\begin{figure}[hbt]
    \centering
    \includegraphics[width=0.6\linewidth]{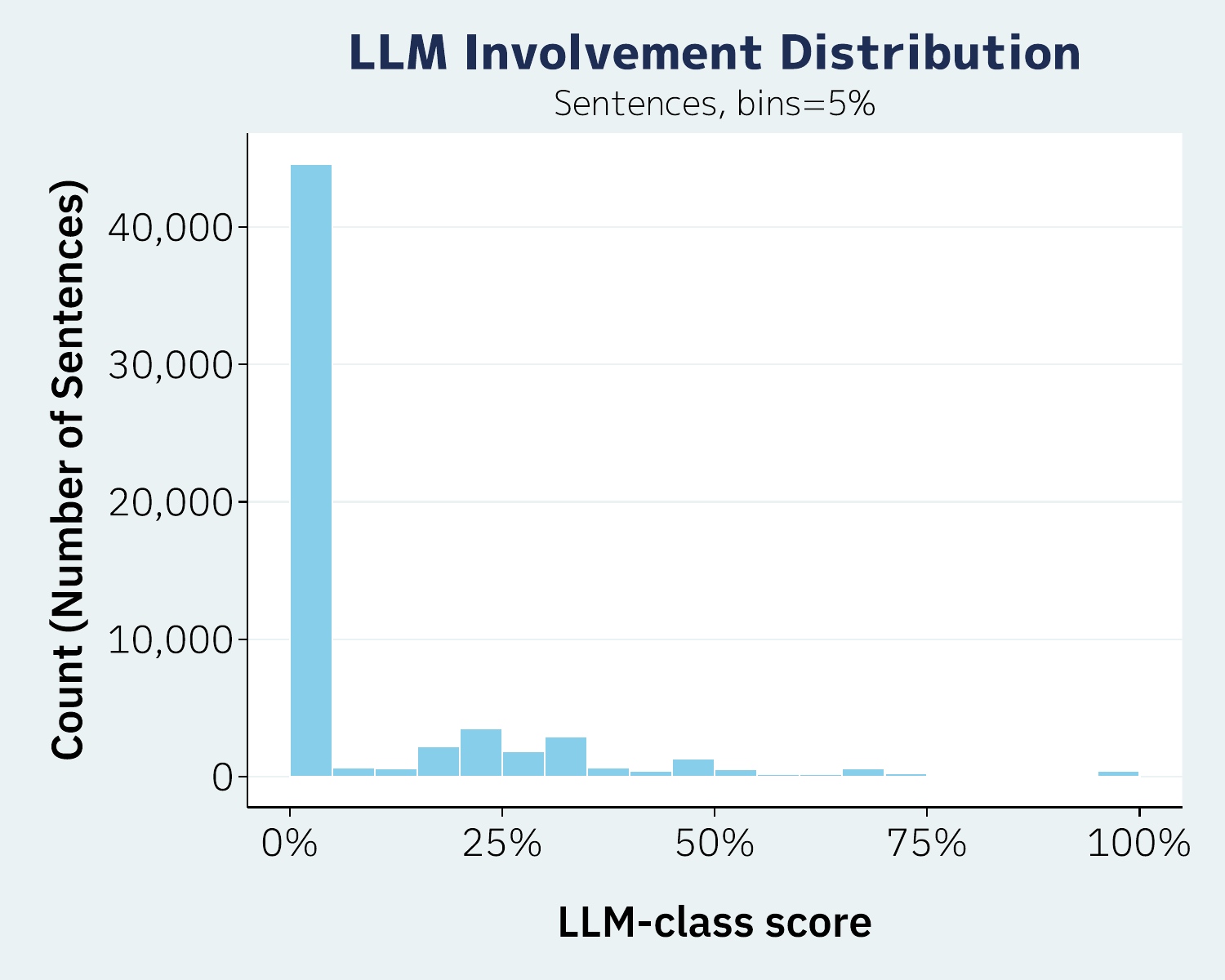}
    \caption{Distribution of estimated AI-generation scores at the sentence level}
    \label{fig:hist_sen}
\end{figure}

For comparison with that study,
the present method was therefore applied at the sentence level to estimate whether individual sentences were AI-generated,
and the resulting score distribution was examined.
The results are shown in \figref{fig:hist_sen}.

As noted above, the results cannot be compared directly.
Nevertheless, the distribution shown in \figref{fig:hist_sen} is somewhat similar in shape to the estimated distribution of AI-generated text in \cite{Qian2026}: it has a peak near 0 and a second, lower and broader mode centered at approximately \qty{25}{\percent} and extending from \qty{0}{\percent} to \qty{50}{\percent}.

%---------------
\section{Trial Estimation Using Word Distributions}
For comparison with previous research, LLM involvement was also estimated as a supplementary analysis using the word-occurrence-distribution and maximum-likelihood method based on \cite{Liang2025}.
For human-written and AI-generated texts, the method estimates the probabilities $p_t$ and $q_t$ that each word occurs in a sentence.
It then assumes that the observed set of sentences was generated from a mixture of the two distributions
and estimates the mixture parameter $\alpha$ by maximum likelihood.

\subsection{Performance Evaluation}

As with the BERT classifier, performance was evaluated using five-fold cross-validation with the original research project ID as the grouping variable.
For each fold, $p_t$ and $q_t$ were estimated using only the training data,
and $\hat{\alpha}$ was calculated for the validation data not used for estimation.

\begin{table*}[htb]
    \centering
    \caption{Five-fold cross-validation results for the MLE method}
    \label{tab:mle_cv}
    \begin{tabular}{lrrrrrrr}
        \toprule
        & \textbf{Accuracy}
        & \multicolumn{3}{c}{\textbf{Macro}}
        & \multicolumn{3}{c}{\textbf{LLM class}} \\
        \cmidrule(lr){3-5}
        \cmidrule(lr){6-8}
        Fold
        &
        & Precision
        & Recall
        & F1
        & Precision
        & Recall
        & F1 \\
        \midrule
        1 & \num{0.9311} & \num{0.9242} & \num{0.9204} & \num{0.9223} & \num{0.9443} & \num{0.9527} & \num{0.9485} \\
        2 & \num{0.9323} & \num{0.9257} & \num{0.9214} & \num{0.9235} & \num{0.9447} & \num{0.9541} & \num{0.9494} \\
        3 & \num{0.9316} & \num{0.9245} & \num{0.9212} & \num{0.9228} & \num{0.9452} & \num{0.9525} & \num{0.9488} \\
        4 & \num{0.9318} & \num{0.9249} & \num{0.9212} & \num{0.9230} & \num{0.9449} & \num{0.9531} & \num{0.9490} \\
        5 & \num{0.9348} & \num{0.9281} & \num{0.9248} & \num{0.9264} & \num{0.9476} & \num{0.9549} & \num{0.9512} \\
        \midrule
        Mean
          & \num{0.9323} & \num{0.9255} & \num{0.9218} & \num{0.9236}
          & \num{0.9453} & \num{0.9534} & \num{0.9494} \\
        SD
          & \num{0.0014} & \num{0.0016} & \num{0.0017} & \num{0.0016}
          & \num{0.0013} & \num{0.0010} & \num{0.0011} \\
        OOF
          & \num{0.9323} & \num{0.9255} & \num{0.9218} & \num{0.9236}
          & \num{0.9453} & \num{0.9534} & \num{0.9494} \\
        \bottomrule
    \end{tabular}
\end{table*}

The OOF estimates produced a mean $\hat{\alpha}$ of \num{0.149} for human-written texts and \num{0.888} for AI-generated texts, a difference of \num{0.740} between the two means.
When human-written texts were assigned a value of \num{0} and AI-generated texts a value of \num{1}, the MAE was \num{0.124}, the RMSE was \num{0.242}, and the ROC-AUC was \num{0.961}.
These results show that the method based solely on word-occurrence distributions distinguished human-written from AI-generated text with high accuracy.
However, even for validation texts known to be entirely human-written or entirely AI-generated, the mean $\hat{\alpha}$ values were \num{0.149} and \num{0.888}, respectively, confirming that the estimates did not necessarily coincide with the true source labels of \num{0} or \num{1}.

For reference,
when the results were converted into binary classifications using $\hat{\alpha}=0.5$ as the threshold, Accuracy was \num{0.932}, Macro F1 was \num{0.924}, and F1 for the LLM class was \num{0.949}
(\tabref{tab:mle_cv})
.
The corresponding values for the BERT classifier used in the main analysis were \num{0.997}, \num{0.997}, and \num{0.998}, respectively, indicating that BERT performed better for the binary distinction between human-written and AI-generated text.

However, the word-distribution method is intended to estimate LLM involvement in a text as a continuous mixture parameter, whereas the BERT classifier is designed for binary classification.
The comparison of their classification performance should therefore be regarded as supplementary.

\subsection{Results for Individual Abstracts}\label{apdx:individual_abstract_results}
The constructed model and parameters were used to estimate the LLM mixture proportion for individual projects from fiscal year 2022 onward.
The results are shown in \tabref{tab:mle_estimation_results}.

\begin{table}[!htbp]
    \centering
    \caption{Estimation results based on MLE}
    \label{tab:mle_estimation_results}
    {
    \begin{tabular}{lp{3mm}rp{1mm}r}
    \toprule
       \textbf{Fiscal Year} && \textbf{Mean ($\hat{\alpha}$)} && \textbf{PCT($\hat{\alpha}\ge 0.5$)} \\
    \midrule
    2022 && \num{0.183} && \qty{13.7}{\percent}\\
    2023 && \num{0.187} && \qty{14.0}{\percent}\\
    2024 && \num{0.192} && \qty{14.5}{\percent}\\
    2025 && \num{0.241} && \qty{20.2}{\percent}\\
    2026 && \num{0.390} && \qty{37.3}{\percent}\\
    \bottomrule
    \end{tabular}
    }
\end{table}

As shown in \tabref{tab:mle_estimation_results}, even in fiscal year 2022 the mean estimate was
$\hat{\alpha}=0.183$,
and \qty{13.7}{\percent} of the abstracts had $\hat{\alpha}\ge 0.5$.
Thus, even in fiscal year 2022, when generative AI use is expected to have been very limited,
\qty{13.7}{\percent} of the abstracts were classified as AI-generated because $\hat{\alpha}\ge0.5$.
Because the mean $\hat{\alpha}$ for human-written texts was also \num{0.149} in cross-validation,
neither the absolute value of $\hat{\alpha}$ nor the proportion exceeding the threshold of \num{0.5} should be interpreted directly as the rate of generative AI use.

Nevertheless, although the absolute levels of the MLE estimates differed from the binary classification results produced by BERT, the temporal pattern was common to both methods: estimates remained broadly stable from 2022 through 2024, began to rise in 2025, and increased substantially in 2026.

The consistency of the temporal trend across two substantially different estimation approaches strengthens confidence that the observed increase is not merely an artifact of a particular modeling framework.

This suggests that the increase observed from 2025 onward is not attributable solely to characteristics specific to the BERT classifier.

\end{document}